\documentclass[pre,aps,showpacs,twocolumn,superscriptaddress]{revtex4}
\usepackage{amsfonts}
\usepackage{amsmath}
\usepackage{amssymb}
\usepackage{graphicx}
\usepackage{txfonts}

\begin{document}

\title{High-precision evaluation of Wigner's d-matrix by exact
diagonalization}
\author{X. M. Feng}
\affiliation{Department of Physics, Beijing Jiaotong University, Beijing 100044, China}
\author{P. Wang}
\affiliation{Beijing Computational Science Research Center, Beijing 100084, China}
\affiliation{Hefei National Laboratory for Physics Sciences at Microscale and Department
of Modern Physics, University of Science and Technology of China, Hefei,
Anhui 230026, China}
\author{W. Yang}
\email{wenyang@csrc.ac.cn}
\affiliation{Beijing Computational Science Research Center, Beijing 100084, China}
\author{G. R. Jin}
\email{grjin@bjtu.edu.cn}
\affiliation{Department of Physics, Beijing Jiaotong University, Beijing 100044, China}

\begin{abstract}
The precise calculations of the Wigner's d-matrix are important in
various research fields. Due to the presence of large numbers, direct calculations of the matrix using the Wigner's formula suffer from loss of precision. We
present a simple method to avoid this problem by expanding the d-matrix into
a complex Fourier series and calculate the Fourier coefficients by exactly
diagonalizing the angular-momentum operator $J_{y}$ in the eigenbasis of $%
J_{z}$. This method allows us to compute the d-matrix and its various
derivatives for spins up to a few thousand. The precision of the d-matrix
from our method is about $10^{-14}$ for spins up to $100$.
\end{abstract}

\pacs{02.60.-x, 03.65.Fd, 02.20.-a, 21.60.-n}
\maketitle

\section{Introduction}

The spin is a fundamental quantum object and an important candidate for
various quantum technologies such as magnetic resonance spectroscopy,
quantum metrology, and quantum information processing. An essential
requirement in these developments is the precise control of many spins or
alternatively a large spin composed of the constituent spins. The simplest
case of such control is the rotation around a fixed axis. Accurately
describing this process requires high-precision calculations of the Wigner's
d-matrix~\cite{Rose,Wigner,Edmonds,Sakurai} that quantifies the rotation of
angle $\theta $ around the $y$ axis: $d_{m,n}^{j}(\theta )\equiv \langle
j,m|\exp(-i\theta J_{y})|j,n\rangle \in \mathbb{R}$, where $|j,m\rangle $ is
an eigenstate of $J_{z}$ with eigenvalue $m$, i.e., $J_{z}|j,m\rangle
=m|j,m\rangle $. Hereinafter, $\hbar =1$ and $i=\sqrt{-1}$.

High-precision calculations of the d-matrix is of interest in quantum
metrology~\cite{Fisher,Helstrom,Pezze}. For instance, let us consider an
atomic Ramsey (or equivalently, Mach-Zehnder) interferometer fed with all
spins down as the paradigmatic setup of interferometric phase estimation.
These spins then undergo an unknown phase shift $\theta$ via the evolution $%
\exp(-i\theta J_{y})$ inside the interferometer. Finally, by detecting the
population imbalance at the output port of the interferometer, i.e., the $%
J_{z}$ measurement with respect to the output state $\exp(-i\theta
J_{y})|j,-j\rangle $, one can record $(2j+1)$ possible outcomes. The outcome
$m$ occurs with probability $P_{m}(\theta )=|\langle j,m|\exp(-i\theta
J_{y})|j,-j\rangle |^{2}=[d_{m,-j}^{j}(\theta )]^{2}$ conditioned on the
unknown parameter $\theta$, thus $\theta$ can be inferred from appropriate
data processing on the outcomes. This process, however, requires accurate
evaluation of Wigner's d-matrix. In addition, the ultimate sensitivity of
this estimation is determined by the Fisher information~\cite%
{Fisher,Helstrom,Pezze}: $F(\theta )\equiv \sum_{m}[\partial P_{m}(\theta
)/\partial \theta ]^{2}/P_{m}(\theta )$, which requires accurate evaluation
of the first-order derivative of Wigner's d-matrix.

In addition to quantum metrology, the Wigner's d-matrix is
closely related to spherical harmonics and Legendre polynomials and is of
interest in many other fields~\cite{Condon,Biedenharn,Varshalovich,Miyazaki,Mueller,Yang,Aubert}. However,
the calculation of the d-matrix for large spins ($j\gg 1$) suffers from a
serious loss of precision, due to the presence of large numbers that exceed
the floating-point precision in Wigner's original formula~\cite{Rose,Wigner,Edmonds,Sakurai}. To avoid this problem, the d-matrix has been
calculated by means of recurrence relations~\cite{Edmonds}. This method
still encounters severe numerical instability in the case of high spin,
although a few remedies have been proposed~\cite{Steinborn,Biedenharn81,Risbo,White,Ivanic,Blanco,Choi,Dachsel,Prezeau,Gumerov}. Recently, Gumerov and Duraiswami~\cite{Gumerov} have developed a new
recursion relation for each subspace of spins, which greatly improves the
stability. However, the achievable
precision (i.e., the maximum absolute error) of their results remains unclear. Most recently, Tajima~\cite%
{Tajima} proposed Fourier-series expansion of the Wigner's d-matrix and
convert the accurate evaluation of the d-matrix to that of the Fourier
coefficients. Such a Fourier-series representation has been shown to be more
useful in improving the numerical stability and the precision. However, each
Fourier coefficient is still a sum of many large numbers that exceed the
floating-point precision, so it has to be evaluated with the assistance of a
formula-manipulation software~\cite{Tajima}.

In this paper, we put forward a very simple method to resolve the above
large-number problem in evaluating the Fourier coefficients of Wigner's
d-matrix~\cite{Tajima}. The essential idea is to express these coefficients
via the inner products $\langle j,m|j, \mu\rangle_y$, where the eigenstates
of $J_{y}$, i.e., $\{|j,\mu\rangle_{y}\}$ constitute an orthonormalized and
completed set. To evaluate such inner products, we write down $J_{y}$ as a
Hermitian matrix in the eigenbasis of $J_{z}$. Then we numerically
diagonalize the $J_{y}$ matrix to obtain the eigenstates and the inner
products. Due to the normalization of $|j,\mu \rangle _{y}$, the norm of
each Fourier coefficient is not larger than unity, thus we avoid the
large-number problem in floating-point calculations. This method allows us
to evaluate accurately the d-matrix and its various derivatives for much
larger spins up to a few thousands, with an absolute error $O(10^{-14})$ for
the d-matrix and $O(j^{k}10^{-14})$ for its $k$th-order derivative.

\section{Fourier series of Wigner's d-matrix}

For large spin $j$, numerical calculation of the Wigner's formula is subject to intolerable numerical errors because it is a sum of
many large numbers with alternating signs~\cite{Rose,Wigner,Edmonds,Sakurai}:
\begin{equation}
d_{m,n}^{j}(\theta )\!=\!\sum_{k}w_{k}^{(j,m,n)}\! \left( \cos {\frac{%
\theta }{2}}\right) ^{2j-2k+n-m}\!\!\left( \sin {\frac{\theta }{2}}\right)
^{2k+m-n},  \label{formula}
\end{equation}%
where
\begin{equation*}
w_{k}^{(j,m,n)}\!=\!\frac{(-1)^{k+m-n}\sqrt{(j+m)!(j-m)!(j+n)!(j-n)!}}{%
(j-m-k)!(j+n-k)!(k+m-n)!k!},
\end{equation*}%
and $k\in \lbrack \max (0,n-m),\min (j-m,j+n)]$. Taking $d_{0,0}^{j}(\pi /2)$ as
an example, the term $k=j/2$ has a very large magnitude $%
|w_{j/2}^{(j,0,0)}|/2^{j}\varpropto 2^{j}/j$, exceeding the floating-point
precision when $j\gg 1$.

To avoid this problem, Tajima~\cite{Tajima} has proposed to expand the
Wigner's d-matrix into a complex Fourier series:
\begin{equation}
d_{m,n}^{j}(\theta )=\sum_{\mu =-j}^{+j}e^{-i\mu \theta }t_{\mu }^{(j,m,n)}.
\label{Fourier}
\end{equation}%
This representation of the d-matrix is very useful and free from the
large-number errors since each Fourier coefficient is less than or equal to
$1$ (see below). However, an accurate evaluation of the Fourier coefficients
$t_{\mu }^{(j,m,n)}$ by means of Eq.~(\ref{formula}) remains nontrivial.
This is because the large-number problem still exists in the coefficients:
\begin{eqnarray}
t_{\mu }^{(j,m,n)} \!&=&\!\frac{1}{2\pi }\int_{0}^{2\pi }d_{m,n}^{j}(\theta )%
\left[ e^{-i\mu \theta }\right] ^{\ast }d\theta  \notag \\
\!&=&\!\sum_{k=\max (0,n-m)}^{\min (j-m,j+n)}\! w_{k}^{(j,m,n)}I_{\mu
}(2j,2k+m-n),  \label{coeff}
\end{eqnarray}%
where we have introduced an integral%
\begin{eqnarray*}
I_{\mu }(2j,\lambda ) \!\!&\equiv &\!\! \frac{1}{2\pi }\int_{0}^{2\pi
}\left( \cos {\frac{\theta }{2}}\right) ^{2j-\lambda }\left( \sin {\frac{%
\theta }{2}}\right) ^{\lambda }e^{i\mu \theta }d\theta  \notag \\
\!&=&\!\!\frac{1}{2^{2j}}\!\!\sum_{l=\max \{0,-j+\mu +\lambda \}}^{\min
\{\lambda ,j+\mu \}}(-1)^{l-\lambda /2}\binom{2j-\lambda }{j+\mu -l}\binom{%
\lambda }{l},
\end{eqnarray*}%
with $\binom{\lambda }{l}=\lambda !/[l!(\lambda -l)!]$. When $j\gg 1$, some
terms in the integral are still huge (e.g., the term $\lambda =j$, $l=0$,
and $\mu =-j/2$). Tajima~\cite{Tajima} bypassed this problem by employing a
symbolic computation software and then reducing the results to
double-precision floating numbers.

\section{Method of exact diagonalization}

\begin{figure}[hptb]
\begin{centering}
\includegraphics[width=1\columnwidth]{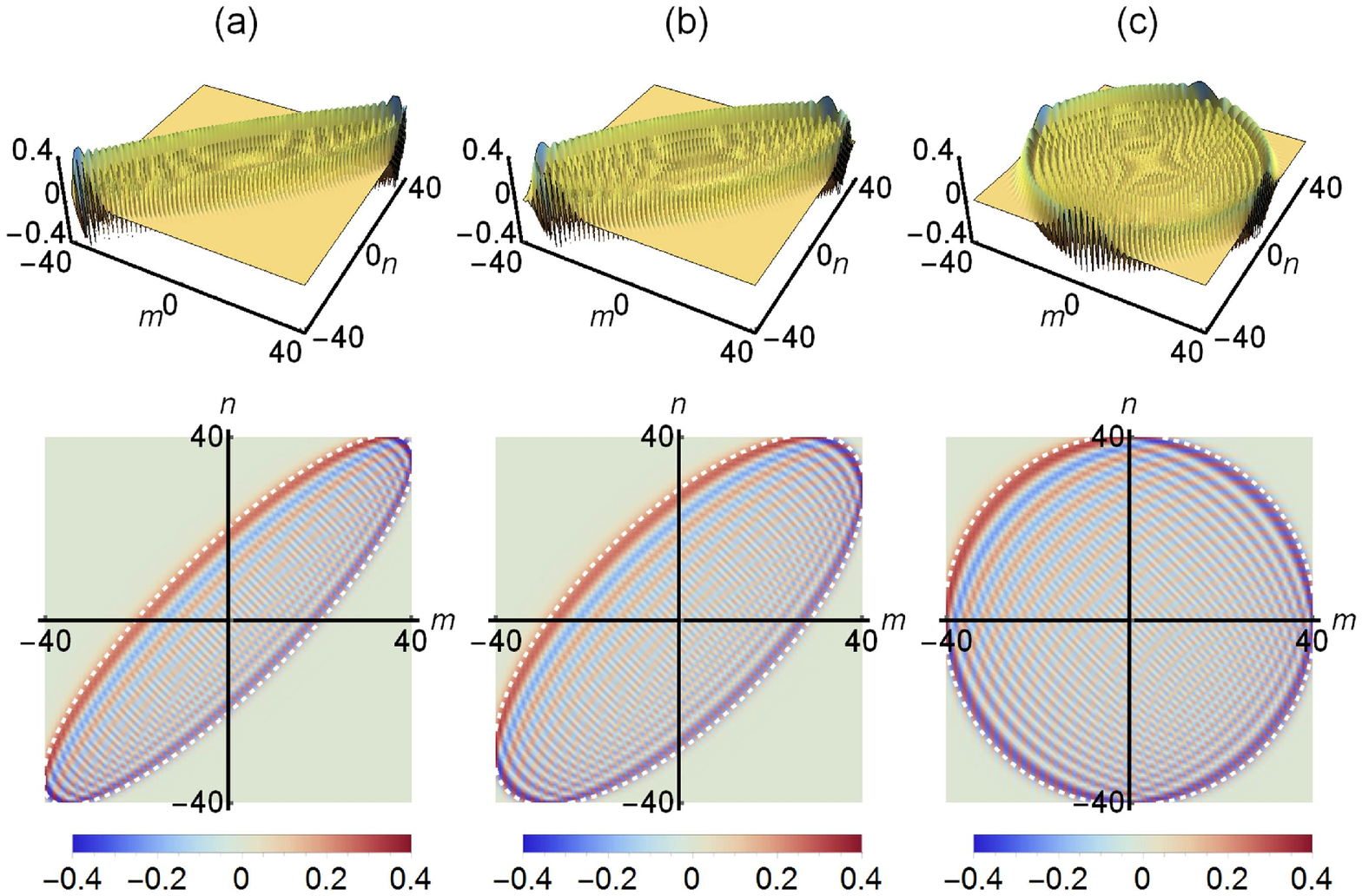}
\caption{(Color online) Computed results of the Wigner's d-matrix $d_{m,n}^{j}(\theta )$ (with total spin $j=40$) against $m$ and $n$ for $\theta=\pi/6$ (a), $\pi/4$ (b), and $\pi/2$ (c), respectively. The dashed lines in bottom panel are the boundary of the central region, determined by Eq.~(\ref{boundary}). Outside the region, the values of $d_{m,n}^{j}(\theta)$ are almost vanishing.}
\label{fig1}
\end{centering}
\end{figure}

Here, instead of using Eq.~(\ref{coeff}), we present a very simple method to
calculate the Fourier coefficients $t_{\mu }^{(j,m,n)}$ that free from the
above mentioned large-number problem. The key observation is that the
d-matrix can be rewritten as
\begin{equation}
d_{m,n}^{j}(\theta )\!\equiv\! \langle j,m|e^{-i\theta
J_{y}}|j,n\rangle \!=\! \sum_{\mu =-j}^{+j}e^{-i\mu \theta }\langle
j,m|j,\mu \rangle _{yy}\langle j,\mu |j,n\rangle ,  \label{dmn3}
\end{equation}%
where $|j,\mu \rangle _{y}\equiv e^{i\frac{\pi }{2}J_{x}}|j,\mu \rangle
=e^{-i\frac{\pi }{2}J_{z}}e^{-i\frac{\pi }{2}J_{y}}e^{i\frac{\pi }{2}%
J_{z}}|j,\mu \rangle $ are eigenstates of $J_{y}$ and they constitute an
ortho-normalized and completed set, i.e., $_{y}\langle j,\mu |j,\mu ^{\prime
}\rangle _{y}=\delta _{\mu ,\mu ^{\prime }}$ and $\sum_{\mu }|j,\mu \rangle
_{yy}\langle j,\mu |=1$. Hereafter, we use $|j,m\rangle $ for the
eigenstates of $J_{z}$ and $|j,\mu \rangle _{y}$ for the eigenstates of $%
J_{y}$. Comparing Eq.~(\ref{Fourier}) and Eq.~(\ref{dmn3}), we identify the
Fourier coefficients in Eq.~(\ref{Fourier}) as
\begin{equation}
t_{\mu }^{(j,m,n)}\!=\! \langle j,m|j,\mu \rangle _{yy}\langle j,\mu
|j,n\rangle \!=\!e^{i\frac{\pi }{2}(n-m)}d_{m,\mu }^{j}\left( \frac{\pi }{2}%
\right) d_{n,\mu }^{j}\left( \frac{\pi }{2}\right) ,  \label{tmn2b}
\end{equation}%
which obeys the sum rule $\sum_{\mu }t_{\mu }^{(j,m,n)}=\langle
j,m|j,n\rangle =\delta _{m,n}$. From Eq.(\ref{tmn2b}), one can note that all
the Fourier coefficients and hence the d-matrix for arbitrary $\theta $
depend on $d_{m,\mu }^{j}(\theta =\pi /2)$ only~\cite{Risbo,Lai,Huffenberger}. The elements of the d-matrix are real and obey the following symmetry properties (see
Refs.~\cite{Rose,Tajima}, and also Fig.~\ref{fig1}):
\begin{eqnarray*}
&&d_{n,m}^{j}( \theta ) =d_{m,n}^{j}(-\theta)=d_{-m,-n}^{j}( \theta )
=(-1)^{n-m}d_{m,n}^{j}( \theta ) , \\
&&d_{m,n}^{j}( \pi -\theta )=(-1)^{j+m}d_{m,-n}^{j}( \theta
)=(-1)^{j-n}d_{-m,n}^{j}(\theta), \\
&&d_{m,n}^{j}(\pi+\theta )=(-1)^{j-n}d_{m, -n}^{j}(\theta )=(-1)^{j+m}d_{-m,
n}^{j}(\theta),
\end{eqnarray*}%
from which one can easily obtain $t_{-\mu }^{(j,m,n)}=(-1)^{2j+m+n}t_{\mu
}^{(j,m,n)}$ and $t_{\mu }^{(j,n,m)}=t_{\mu }^{(j,-m,-n)}=(-1)^{n-m}t_{\mu
}^{(j,m,n)}$, as observed recently by Tajima~\cite{Tajima}.

Most importantly, the first result of Eq.~(\ref{tmn2b}) provides a very
simple but accurate method to calculate the Fourier coefficients and hence
the Wigner's d-matrix by solving the inner product $\langle j,m|j,\mu
\rangle _{y}$. To this end, we first express $J_{y}$ as a ($2j+1$%
)-dimensional Hermitian matrix:%
\begin{equation}
J_{y}=\frac{1}{2i}\left(
\begin{array}{cccccc}
0 & -X_{-j+1} &  &  &  &  \\
X_{j} & 0 & -X_{-j+2} &  &  &  \\
& X_{j-1} & 0 & \ddots &  &  \\
&  & \ddots & \ddots & \ddots &  \\
&  &  & \ddots & 0 & -X_{j} \\
&  &  &  & X_{-j+1} & 0%
\end{array}%
\right) ,  \label{Jy}
\end{equation}%
where the matrix elements are determined by $\langle j,m|J_{y}|j,n\rangle
=(X_{-n}\delta _{m,n+1}-X_{n}\delta _{m,n-1})/(2i)$, with the term $X_{m}=%
\sqrt{(j+m)(j-m+1)}$ satisfying $X_{\pm m}=X_{\mp m+1}$ and $X_{-j}=0$.
Next, we diagonalize the Hermitian matrix using standard numerical methods,
e.g., the ZHBEV subroutine of LAPACK, or the EVCHF package of the IMSL, to
obtain all the eigenvectors $\{|j,\mu \rangle _{y}\}$ and their probability
amplitudes $\langle j,m|j,\mu \rangle _{y}$. For the simplest case $j=1/2$,
the matrix is indeed the transpose of the Pauli matrix $\sigma _{y}$ over
two (i.e., $\sigma _{y}^{T}/2$), which gives two eigenvectors $|1/2,\pm
1/2\rangle _{y}$, corresponding to the eigenvalues $\pm 1/2$. The inner
products $\langle j,m|j,\mu \rangle _{y}$ for $j=1/2$ are given by $\langle
1/2,1/2|1/2,\pm 1/2\rangle _{y}=1/\sqrt{2}$ and $\langle 1/2,-1/2|1/2,\pm
1/2\rangle _{y}=\pm i/\sqrt{2} $. Inserting them into Eq.~(\ref{dmn3}), one
can obtain the Wigner's d-matrix.

The exact-diagonalization method has two advantages. First, the magnitude of
$\langle j,m|j,\mu \rangle _{y}$ and hence all the coefficients $t_{\mu
}^{(j,m,n)}$ in Eq.~(\ref{tmn2b}) are not larger than unity, since all the
eigenvectors $\{|j,\mu \rangle _{y}\}$ are normalized. This provides a
solution to the large-number problem in Eqs.~(\ref{formula}) and (\ref{coeff}%
). Second, the matrix in Eq.~(\ref{Jy}) is tridiagonal and Hermitian, which
can be easily diagonalized. By diagonalizing the $J_y$ matrix only once, all
the Fourier coefficients and hence all the elements of the d-matrix can be
obtained with given $j$ and $\theta$ (see the Supplemental Material~\cite%
{Supplemental}). This method allows us to calculate Wigner's d-matrix for $j$
up to a few thousands. More important, one can compute arbitrary $k$-th
order derivative of the d-matrix with little additional cost, due to
\begin{equation}
\frac{\partial ^{k}d_{m,n}^{j}(\theta )}{\partial \theta ^{k}}\!=\!\langle
j,m| e^{-i\theta J_{y}}(-iJ_{y})^{k}|j,n\rangle \!=\!\sum_{\mu
=-j}^{+j}(-i\mu )^{k}e^{-i\mu \theta }t_{\mu }^{(j,m,n)}.  \label{derivative}
\end{equation}%
This is indeed a common advantage of the Fourier-series representation. As
comparison, one can find that the direct evaluation of the first result of
Eq.~(\ref{derivative}) costs double even for the case $k=1$ since $\langle
j,m|\exp(-i\theta J_{y})(-iJ_{y})|j,n\rangle$ gives
\begin{equation}
\frac{\partial d_{m,n}^{j}(\theta )}{\partial \theta }=\frac{1}{2}\left[%
X_{n}d_{m,n-1}^{j}(\theta )-X_{-n}d_{m,n+1}^{j}(\theta )\right],
\label{1thderivative}
\end{equation}%
which depends on two elements of the d-matrix. When $n=\pm j$, only one of
them remains.

\section{Numerical results and error analysis}

\begin{figure}[hptb]
\begin{centering}
\includegraphics[width=1\columnwidth]{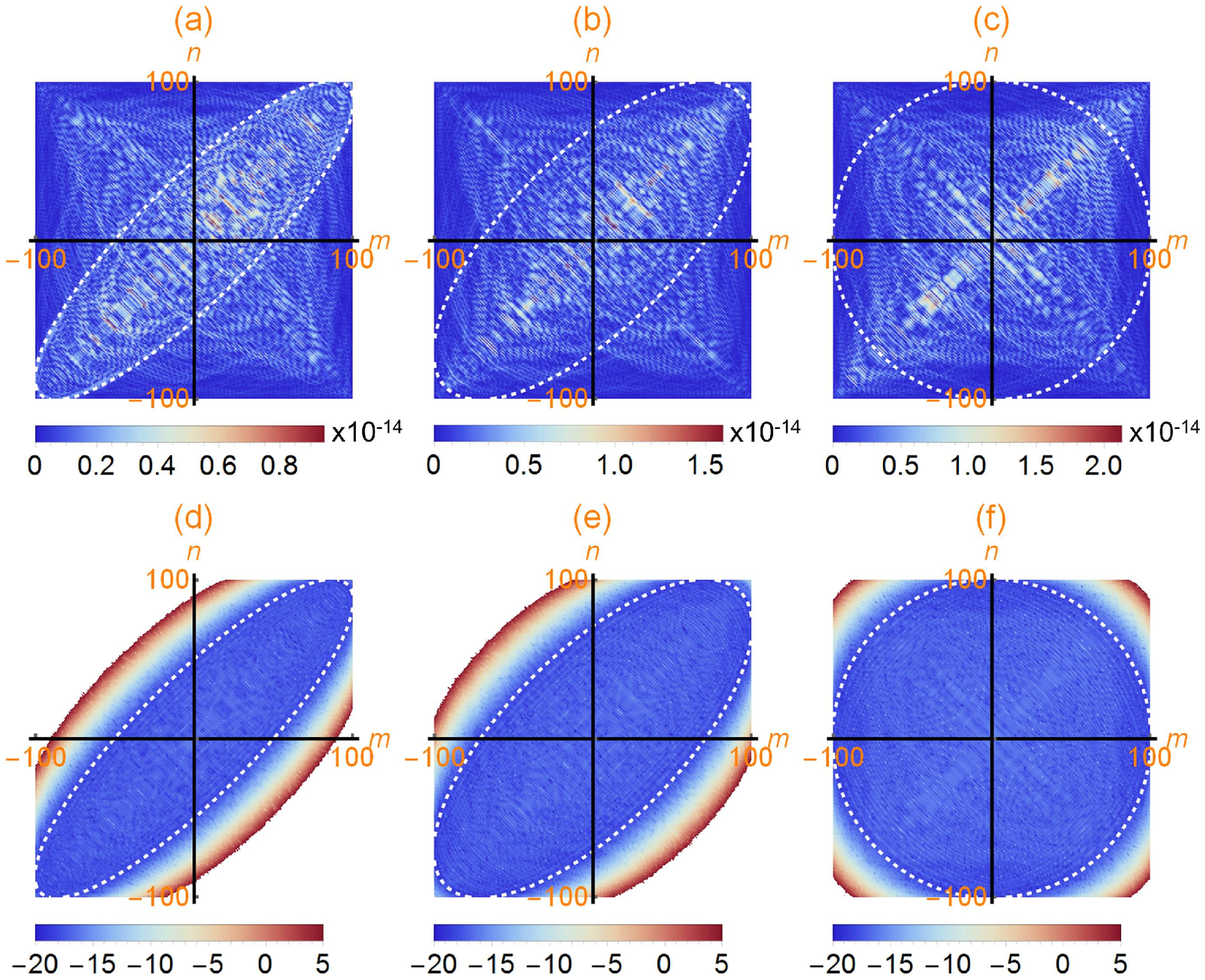}
\caption{(Color online) The absolute error $\Delta _{\mathrm{abs}}=|(d_{\mathrm{comp}}-d_{\mathrm{ex}})|$ (the upper panel) and the relative error $\log _{10}\Delta _{\mathrm{rel}}$ (the lower panel) of the Wigner's d-matrix $d_{m,n}^{j}(\theta)$ with the spin $j=100$, and the rotation angle $\theta=\pi/6$ (left), $\pi/4$ (middle), $\pi/2$ (right), respectively. The computed results are obtained by diagonazing the matrix in Eq.~(\ref{Jy}), using the ZHBEV subroutine of LAPACK; While the exact results $d_{ex}$ are obtained by MATHEMATICA 10.0. The dashed lines are given by Eq.~(\ref{boundary}).}
\label{fig2}
\end{centering}
\end{figure}

As shown in Fig.~\ref{fig1}, we plot the computed results of $%
d_{m,n}^{j}(\theta )$ in the plane ($m,n$), with $m,n\in \lbrack -j,+j]$.
For a relatively large spin $j=40$ and a given $\theta $, $%
d_{m,n}^{j}(\theta )$ is appreciable only in the central region and tend to
zero quickly outside this region. The boundary of this region is determined
by (see also the dashed lines of Fig.~\ref{fig2})
\begin{equation}
m^{2}+n^{2}-2mn\cos \theta =j(j+1)\sin ^{2}\theta ,  \label{boundary}
\end{equation}%
at which $\partial ^{k}d_{m,n}^{j}(\theta )/\partial \theta ^{k}=0$ for $k=1$%
, $2$. This boundary equation follows from the differential equation of the
d-matrix~\cite{Edmonds}. Similar boundary equation has been obtained using
the WKB approximation~\cite{Braun}.

\begin{figure}[hptb]
\begin{centering}
\includegraphics[width=0.95\columnwidth]{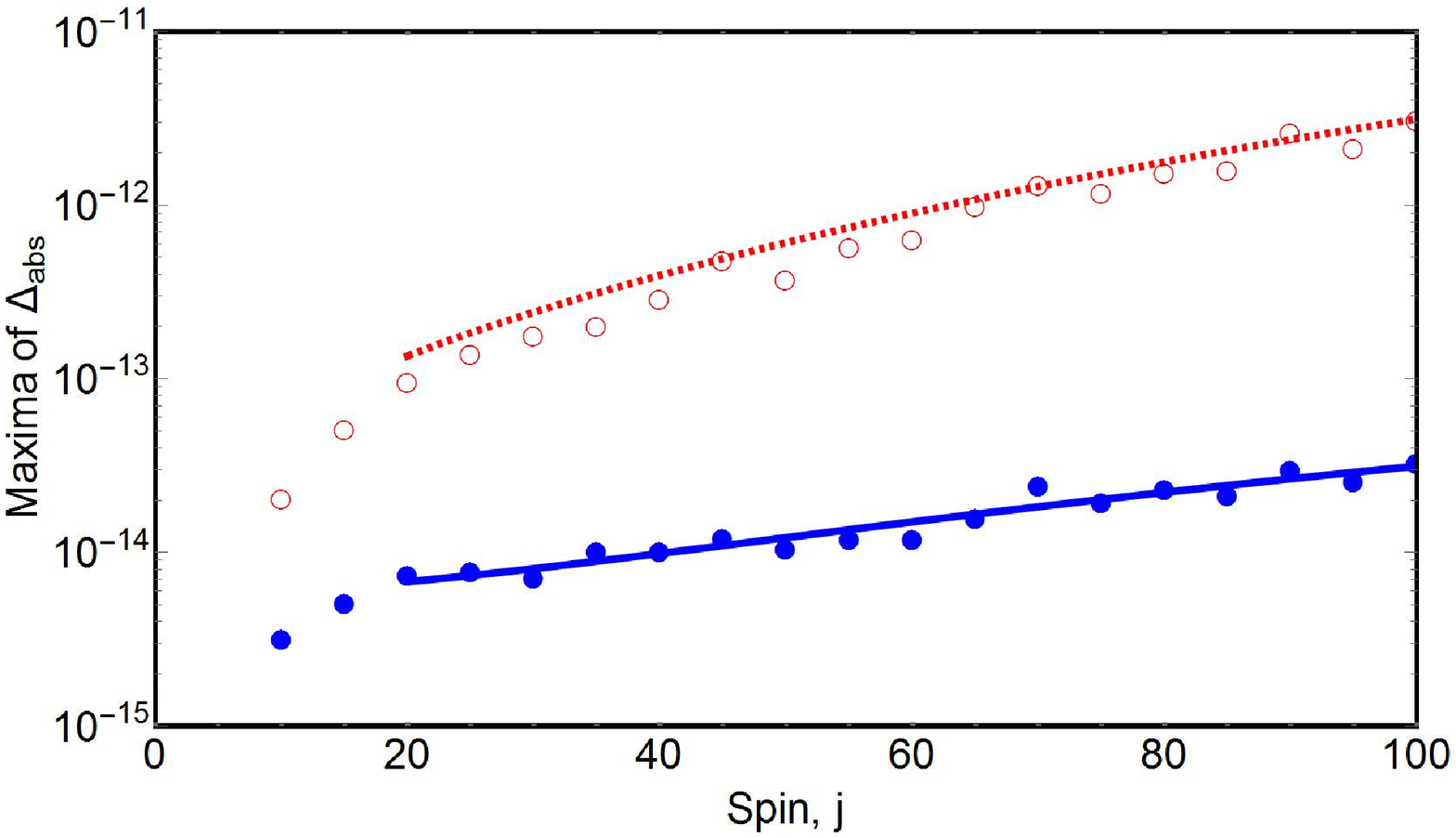}
\caption{(Color online) The maximum absolute error of $d_{m,m}^{j}(\theta)$ (solid circles) and that of $\partial d_{m,m}^{j}(\theta )/\partial\theta $ (open circles) as functions of spin $j$. Blue solid line: the fitting result, $\Delta _{\mathrm{abs,\max}}\approx (a j^{2}+b)\times 10^{-14}$ with $a=2.568\times 10^{-4}$ and $b=0.5758$. The precision of the d-matrix can reach $3.275\times 10^{-14}$ at $j=100$. Red dashed line: maximum error of the first-order derivative $\sim j\times\Delta _{\mathrm{abs,\max}}$. }
\label{fig3}
\end{centering}
\end{figure}

Given the exact value $d_{\mathrm{ex}}$ and the numerically calculated value
$d_{\mathrm{comp}}$ of a matrix element $d_{m,n}^{j}(\theta )$, the absolute
error is defined as $\Delta _{\mathrm{abs}}=|d_{\mathrm{comp}}-d_{\mathrm{ex}%
}|$ and the relative error is defined as $\Delta _{\mathrm{rel}}=|(d_{%
\mathrm{comp}}-d_{\mathrm{ex}})/d_{\mathrm{ex}}|$. The exact values of the
d-matrix elements are obtained by MATHEMATICA 10.0. Figure~\ref{fig2} shows
that the absolute error $\Delta _{\mathrm{abs}}\sim 10^{-14}$ even for a
relatively larger spin $j=100$ (see also Fig.~\ref{fig3}). The lower panel of
Fig.~\ref{fig2} shows the relative error $\lesssim 10^{-10}$ within the
central region (enclosed by the dashed line), but increases rapidly outside
this region.

Outside the boundary, the large relative error simply follows from the very
small exact values $|d_{\mathrm{ex}}|$. To illustrate this point, we
consider $d_{m,n}^{j}(\theta )$ with $|m|=|n|=+j$. In this case, we have an
analytical expression
\begin{equation*}
d_{j,m}^{j}(\theta )=(-1)^{j-m}\binom{2j}{j+m}^{1/2}\left( \cos {\frac{%
\theta }{2}}\right) ^{j+m}\left( \sin {\frac{\theta }{2}}\right) ^{j-m}.
\end{equation*}%
Using the symmetry, one can obtain $d_{\pm j,\mp j}^{j}(\theta )=[-\sin
(\theta /2)]^{2j}$. For $j=100$ and $\theta =\pi /6$, one can easily obtain
the exact values $d_{\pm j,\mp j}^{j}(\pi /6)=3.974\times 10^{-118}$, which
lie outside the boundary [see the dashed line of Fig.~\ref{fig2}(d)]. By
contrast, although the numerically calculated values $d_{\pm j,\mp
j}^{j}(\pi /6)\sim 10^{-17}$ are very close to zero, they are much larger
than the exact values, leading to a large relative error $\Delta _{\mathrm{%
rel}}$.

Due to the same reason, the d-matrix elements for other values of $\theta $
also show large $\Delta _{\mathrm{rel}}$ outside the central region. For
example, $\theta=\pi /2$, the boundary of the central region is a circle: $%
m^{2}+n^{2}=j(j+1) $ with a radius $\sim j$. The d-matrix elements with $%
|m|=|n|=j$ have the same exact value $d_{m,n}^{j}(\pi /2)=1/2^{j}\sim
7.9\times 10^{-31}$, which is much smaller than the numerically calculated
value $\sim 10^{-15}$, yielding a large $\Delta _{\mathrm{rel}}$. The
computed $d_{m,n}^{j}(\pi /2)$ at $mn=0$ also show large relative error. To
see it clearly, we use the exact result~\cite{Rose}
\begin{equation*}
d_{m,0}^{j}(\theta )=(-1)^{m}\sqrt{\frac{(j-m)!}{(j+m)!}}P_{j}^{m}(\cos
\theta ),
\end{equation*}%
where $P_{j}^{m}(x)$ is the associated Legendre polynomial. When $\theta
=\pi /2$, we obtain the exact results $d_{m,0}^{j}(\pi
/2)=(-1)^{m}d_{0,m}^{j}(\pi /2)\propto P_{j}^{m}(0)=0$ for odd $j-m$. In
contrast, the computed results are small but nonzero, thus $\Delta _{\mathrm{%
rel}}\rightarrow \infty $.

Finally, we discuss the scaling of the error in evaluating the d-matrix and
its various derivatives with increasing spin $j$. For this purpose, we sweep
$m,n,$ and $\theta $ and calculate the maximum absolute errors $\Delta _{%
\mathrm{abs,\max }}$ for $d_{m,n}^{j}(\theta )$ and $\partial
d_{m,n}^{j}(\theta )/\partial \theta $ as functions of $j$ for $j$ up to
100. Since the maximum absolute error of the d-matrix almost appears inside
the boundary (see the upper panel of Fig.~\ref{fig2}), we only sweep $(m,n)$
within the central region and increase $\theta$ from $0$ to $\pi /2$ with an
increment $\pi /36$~\cite{Tajima}. We find that typically the maximum
absolute error appears at $m\sim n$ and $\theta \sim 0$ or $\pi /2$. As
shown in Fig.~\ref{fig3}, one can find that numerical results of $%
10^{14}\times \Delta _{\mathrm{abs,\max }}$ can be well fitted by $aj^{2}+b$%
, with $a\sim 10^{-4}$ and $b\sim 0.6$ (see the solid line). This precision
is slightly worse than the previous one $\Delta_{\mathrm{abs,\max }}\approx
10^{-14.8+0.006j}$ for the spin $j$ up to $100$~\cite{Tajima}. However, if
the scaling persists to larger $j$, our method could provide a smaller error
as $j>405$. One can also note that the maximum absolute error of $\partial
d_{m,n}^{j}(\theta )/\partial \theta $ can be approximated by $j\times
\Delta _{\mathrm{abs,\max }}$ (the dashed line). More generally, from Eqs.~(%
\ref{dmn3}) and (\ref{derivative}), we can deduce that the maximum absolute
error in evaluating the $k$th-order derivative $\partial
^{k}d_{m,n}^{j}(\theta )/\partial \theta ^{k}$ is larger
than that of the d-matrix by a factor $O(j^{k})$.

\section{Conclusion}

In summary, we have presented a very simple method to evaluate accurately
the Wigner's d-matrix by diagonalizing the angular-momentum operator $J_{y}$
in the eigenbasis of $J_{z}$. The coefficients of Fourier-series expansion
of the d-matrix, closely related to the eigenstates of $J_{y}$, are shown to
be not larger than unity. This enable us to avoid the large-number problem.
The absolute error of $d_{m,n}^{j}(\theta )$ can reach $\sim 10^{-14}$ for
spin $j$ up to $100$ and the relative error $\sim 10^{-10}$ within the
central region $m^{2}+n^{2}-2mn\cos \theta \leq j(j+1)\sin ^{2}\theta $,
outside which the values of the d-matrix tends to zero quickly. As one of
the main advantages, we show that for given $j$ and $\theta$, all the
elements of Wigner's d-matrix and their $k$-th order derivatives can be
obtained by diagonalizing $J_{y}$ only once~\cite{Supplemental}. The method
presented here is a part of the larger class of algorithms based on the
matrix exponents~\cite{Moler}. As the matrix exponent of $J_y$, the Wigner's
d-matrix could be computed even more accurately and efficiently since the
matrix in Eq.~(\ref{Jy}) is Hermitian and tridiagonal.

\begin{acknowledgments}
The first two authors (X.M.F. and P. W.) contributed equally to this work.
We thank Professor N. Tajima for helpful discussions. This work was
supported by the NSFC (Contracts No.~11174028, No.~11274036, and
No.~11322542), the MOST (Contract No.~2013CB922004, No.~2014CB848700).
\end{acknowledgments}

\end{document}